\documentclass[12pt]{article}
\usepackage{amsmath}
\usepackage{amssymb}

\date{}

\title{Detection model based on representation of quantum particles
by classical random fields: Born's rule and beyond}

\author{Andrei Khrennikov\\International Center for Mathematical Modelling
\\in Physics and Cognitive Sciences,\\
University of V\"axj\"o, S-35195, Sweden\\
Andrei.Khrennikov@vxu.se}

\begin{document}

\maketitle

\abstract{Recently a new attempt to go beyond quantum mechanics
(QM) was presented in the form of so called prequantum classical
statistical field theory (PCSFT).  Its main experimental
prediction is violation of Born's rule which provides only an
approximative description of real probabilities. We expect that it
will be possible to design numerous experiments demonstrating
violation of Born's rule. Moreover, recently the first
experimental evidence of violation was found in the triple slits
interference experiment, see \cite{WWW}. Although this
experimental test was motivated by another prequantum model,  it
can  be definitely considered as at least preliminary confirmation
of the main prediction of PCSFT. In our approach quantum particles
are just symbolic representations of ``prequantum random fields,''
e.g., ``electron-field'' or ``neutron-field''; photon is
associated with classical random electromagnetic field. Such
prequantum fields fluctuate on time and space scales which are
essentially finer than scales of QM, cf. `t Hooft's attempt to go
beyond QM \cite{H1}--\cite{TH2}. In this paper we elaborate a
detection model in the PCSFT-framework. In this  model classical
random fields (corresponding to ``quantum particles'') interact
with detectors inducing probabilities which match with Born's rule
only approximately. Thus QM arises from PCSFT as an approximative
theory. New tests of violation of Born's rule are proposed.}

\section{Introduction}

Since the first days of creation of QM, its formalism and especially its interpretation
have been inducing permanent debates. Bohr's postulate on completeness of QM, i.e., impossibility to go beyond it and to develop
a finer description of micro-world than given by QM-formalism,
was especially stimulating for these debates -- starting with work of Einstein-Podolsky-Rosen
\cite{E}.  In spite of
so called no-go theorems  -- von Neumann, Bell, Kochen-Specker,... (see \cite{DEM4}, \cite{KB} or
\cite{A} for discussions)
various attempts to go beyond QM have been done during last
80 years. We can mention De Broglies' theory of double solution which was later elaborated
in Bohmian mechanics, stochastic electrodynamics (SED), semiclassical model for quantum
optics, Nelson's stochastic QM and its generalization by Davidson and, recently,
`t Hooft's model, see, e.g., \cite{R1}--\cite{RF},
\cite{H1}, \cite{H2}, also cf. V. I. Manko and O. V. Manko \cite{M1}, \cite{M2} as well
as  Bracken and Wood \cite{B1}, \cite{B2}; ; see also recent paper of T. Else on justification of
`t Hooft's approach \cite{TE}. These models have their own advantages and
disadvantages. In any event, their creation demonstrated that some possibilities to go beyond
QM still can be found (in spite no-go theorems).

Recently \cite{KHR} -- \cite{KHR2} a new attempt to
go beyond quantum mechanics (QM) was done in the form
of so called prequantum classical statistical field theory (PCSFT).
Hidden variables are given by classical fields. These prequantum fields fluctuate on time and space scales
which are essentially finer than scales of QM, cf. `t Hooft's attempt to go beyond QM
\cite{H1}, \cite{H2}. However, opposite to Einstein, PCSFT does not assign results of quantum
observations directly to hidden variables. Bohr's idea that the whole experimental
arrangement should be taken into account is basic for PCSFT-based measurement
theory, see De Muynck \cite{DM}  or D´Ariano \cite {DARIANO}
for the modern presentation.
This theory for measurements of positions of quantum particles as well as discrete
observables  is presented
in this paper. Thus our theory {\it combines peacefully ideas of both Einstein and Bohr.}
On the one hand, we deny Bohr's philosophic principle on completeness of QM.
On the other hand, we deny naive Einsteinian realism -- assigning results of all
possible observations to a hidden variable.

We remind that the {\it semiclassical model} provides a
simple formalism for calculation of probabilities of photon
detection, see, e.g., \cite{R1}--\cite{R2D}. The semiclassical model
is a purely classical field model. In particular, such a
fundamental of the quantum mechanical description as
noncommutativity is eliminated. Quantum observables are presented
by classical random variables. However, the semiclassical model is
applicable only to the electromagnetic field. Moreover, not all
probability distributions of QM can be reproduced, see, e.g., Scully
and Zubairy \cite{R1} for discussions.

We shall also consider a model with prequantum field variables.
Comparing the semiclassical model with our model (which we call
{\it prequantum classical statistical field theory} - PCSFT), we
can say:

\medskip

a)  PCSFT is based on a different mathematical model of
random field;

\medskip

b) it is applicable not only to the electromagnetic
field, but for any type of field;

\medskip

c) it reproduces Born's rule in the general framework, hence, it
reproduces any quantum probability distribution.

\medskip

d) Born's rule appears as an approximative formula for calculation
of real probabilities; hence, it can be violated in better
designed experiments.

\medskip

Regarding b): in PCSFT the basic objects
are not particles, but classical fields, by b) any type of
quantum particle is described by the corresponding prequantum
random field. For example, we can consider the electron-field  or
the neutron-field.

The foundations of PCSFT were given in author's papers
\cite{KHR}-- \cite{KHR2}. It was shown that quantum
average\footnote{It is given by von Neumann's trace formula which
is  based on Born's rule.} gives the main contribution to
corresponding average with respect  to classical prequantum random
field. However, in general quantum and prequantum averages do not
coincide. If one believe that nature is correctly described by
PCSFT and not by QM (and the latter is just an {\it approximative
model}), then one can expect to find experimental deviations from
predictions of QM.

\medskip

However, measurement theory based on PCSFT
has not been developed in \cite{KHR}-- \cite{KHR2}.
The absence of  the corresponding  measurement theory induced the impression
that  PCSFT is a kind  of ontic theory which does not have a direct
relation with quantum measurements. In the present paper
measurement theory based on  PCSFT will be presented. The basic
idea is that a so called quantum state (e.g., a pure state) is
simply a label for an ensemble of classical fields -- {\it classical
random field.} We describe the process of interaction of  detectors
with a random field.

First we consider the simplest model
of measurement in which  only the square-type nonlinearities of
the prequantum field,  $\vert \phi(x) \vert^2,$
are taken into account.\footnote{In PCSFT average of any prequantum random
field $\phi(x, \omega),$ where $\omega$ is a random parameter, is
 the ``zero field'':  $\phi_0(x) = 0$  for any $x \in {\bf R}^3.$
Dispersion of $\phi(x, \omega)$ is very small. Therefore for
``statistical majority'' of realizations of $\phi(x, \omega)$ its amplitude is very small.
Thus the term based on $\vert \phi(x) \vert^2$ gives the main contribution into detector's
output. The contributions of higher order nonlinearities,  $\vert \phi(x) \vert^n, n> 2,$ are essentially smaller. The first
approximation of detection is based on taking into account only quadratic nonlinearity
$\vert \phi(x) \vert^2.$ }  In such a model \cite{PLA} the probability of
selection of a field  is proportional to its ``power''  -- its
$L_2$-norm. We call this approach (quadratic)  {\it power
signal-field detection theory - PFSDT.} The power of signal (classical
prequantum field)  plays the crucial role
in PFSDT. It will be shown that in such a detection model {\it Born's rule} is
a consequence of the well known {\it Bayes formula} for
conditional probability.\footnote{Here everything is classical -- fields,
detectors, nevertheless, the output probabilities are the same as
in QM.}

By considering experiments in which the contribution of
higher powers of the prequantum field can be taken into account,
one can expect {\it deviation from
Born's rule.} Our calculations, see section \ref{GGG}, provide an estimate for such
deviation. Such a detection theory we will also call PFSDT (with in general nonquadratic power).

Recently the first experimental evidence of violation of Born's was found
in the triple slits interference experiment, see \cite{WWW}. It provides at least preliminary
confirmation of the main prediction of PCSFT, see section \ref{SOR} for discussion.

We remind that the dimension
of the space of classical fields is infinite. This induces essential mathematical
difficulties. To escape these difficulties and at the same time to illustrate
all distinguishing features of our model of detection of classical fields-signals,
we start with a toy model with finite-dimensional state space.

\section{Finite-dimensional model}

Let  us consider systems -- ``fields'' -- with the state space ${\cal H}_n, $ where
${\cal H}_n$ is the $n$-dimensional real space: ${\cal H}_n={\bf R}
\times \ldots \times {\bf R}.$ In this model  the  state of a system is given by a vector
$v=(v_1, \ldots, v_n).$ We set $||v||^2=\sum_{j=1}^n v_j^2.$
It will be fruitful to use the analogy with classical signal theory. We shall also call
states of systems {\it signals.} The Euclidean norm of a signal, $||v||^2,$ we shall call
{\it signal power.}

Suppose we have a measurement device ("detector")  that produces one of the
values $j =1,..., n:$
\begin{equation}
\label{J}
j=j(v)
\end{equation}
for a system having the state $v.$

Suppose also that the probability to obtain the fixed value $j_0$ is proportional to
$v_{j_0}^2$ and the coefficient of proportionality does not depend on
$j_0$ (but it depends on the state $v).$  Thus the probability that such a
 detector produces the result
$j=j_0$ for a system with state $v$ is given by
\begin{equation}
\label{J1}
P (j=j_0|v)=k_v v_{j_0}^2.
\end{equation}
The coefficient $k_v$ can be found from the normalization of
probability by one:
\begin{equation}
\label{JU}
\sum_{j_0=1}^n P(j=j_0|v)= k_v \sum_{j_0=1}^n v_{j_0}^2=1.
\end{equation}
Thus $k_v=\frac{1}{||v||^2}$ and
\begin{equation}
\label{J1T}
P(j=j_0|v)= \frac{v_{j_0}^2}{||v||^2}.
\end{equation}

Consider now {\it complex systems combined of  ensembles of ``fields''.}
To keep closer to the analogy with QM, we can call such complex systems
{\it ``particles.''}  Mathematically  a ``particle''  is  described by a
probability measure $\mu$ on ${\cal H}_n,$ or by a random vector
$$\xi(\omega)=(\xi_1(\omega), \ldots, \xi_n(\omega)).$$ It is
assumed that $m_\mu=\int_{{\cal H}_n} vd\mu (v)=0$ (zero mean
value).  Its dispersion
$$
\sigma^2(\mu)=\int_{{\cal H}_n}||v||^2 d\mu (v)=\kappa,
$$
will play an important role in our considerations.

We consider the following procedure of measurement of
the quantity $j$ for  ``particles'':

a) first the detector selects a fixed ``field'' $v$  from the ensemble representing
a particle;

b) then the detector performs measurement of $j(v)$ for this selected $v$
(by the rule which has been already formulated).

Consider a  {\it special  procedure of  selection} of a fixed ``field'' $v$  from the
ensemble (realization of the random vector).  Our basic assumption is that a measurement
device under consideration works in the following way. The probability to
select a ``field''  with the state $v$  from the ensemble is proportional to the {\it square
of the norm of the state}:
$$
P_\mu(v) = K_\mu\; ||v||^2= K_\mu\; \sum_{j=1}^n v_j^2.
$$
The coefficient of proportionality can be again found from the
normalization of probability by one:
$$
1= \int_{{\cal H}_n} d P_\mu (v)=K_\mu \int_{{\cal
H}_n}||v||^2 d\mu (v).
$$
Thus
\begin{equation}
\label{GL} K_\mu =\frac{1}{\kappa}
\end{equation}
and
\begin{equation}
\label{GL2} d P_\mu(v)=\frac{|| v ||^2} {\kappa}d\mu (v).
\end{equation}

Let $\mu$ be a discrete probability measure. It is concentrated in
a finite number of points $v^{(1)},..., v^{(m)}$ of the state space.
It is determined by positive  weights $\mu (v^{k}), k=1,.., m,$
where
$$
\sum_{k=1}^m \mu (v^{k})= 1, v^{(1)}, \ldots, v^{(m)} \in {\cal H}_n.
$$
Now the formula (\ref{GL2}) takes the form:
\begin{equation}
\label{GL2U} P_\mu(v^{(i)})=\frac{||v^{(i)}||^2}{\kappa}
\mu(v^{(i)}).
\end{equation}
It gives the probability that a measurement device (of the class under consideration)
would select (for the subsequent $j$-measurement) a ``field''
having the state $v^{(i)}=(v_1^{(i)},..., v_n^{(i)}).$

In the discrete case we now find  the final probability to obtain
the concrete result $j=j_0: {\bf p}_\mu(j=j_0).$
Let us apply the Bayes  formula.   Probability ${\bf p}_\mu(j=j_0)$
is produced as the result of combination (with the aid of the
Bayes rule)  of  two  probabilities:  probability
$P_\mu(v^{(i)})$ to select a ``field'' with the state $v= v^{(i)}$
from the ensemble of  ``fields'' composing a ``particle'' and
probability $P (j=j_0|v^{(i)})$ to obtain the result $j=j_0$ for
measurement on a  ``field''  with the  state $v^{(i)}:$
\begin{equation}
\label{GL2P} {\bf p}_\mu(j=j_0)= \sum_i P_\mu(v^{(i)}) \; P (j=j_0|v^{(i)}).
\end{equation}
We put in this formula the probabilities given by (\ref{GL2U}) and (\ref{J1T}) and obtain:
\begin{equation}
\label{GL2W} {\bf p}_\mu(j=j_0) = \frac{1}{\kappa} \sum_i \vert v_{j_0}^{(i)}\vert^2
\mu(v^{(i)}).
\end{equation}

\medskip

We now generalize (\ref{GL2W} ) to the case of continuous
distribution $\mu.$ We have: $
P(j=j_0|v)=\frac{v_{j_0}^2}{||v||^2} $ and $
P(v)=\frac{||v||^2}{\kappa} \mu (v). $ Again by using the Bayes
formula we find the probability to obtain $j=j_0$ for measurement
on a system from the ensemble given by $\mu:$
$$
{\bf p}_\mu(j=j_0)=\int_{{\cal H}_n} P (j=j_0|v) d P_\mu(v)=
\frac{1}{\kappa} \int v_{j_0}^2 d\mu (v).
$$

\subsection{Pure states}
\label{TTT}

Let take a vector  $u \in {\cal H}_n$ such that  $||u||=\kappa.$
Consider now a Gaussian measure $\mu \equiv \mu_u$ having the zero
mean value and the covariance operator
$$
C_u=u \otimes u,
$$
i.e., $(C_u y, y) = (u, y)^2.$ This Gaussian measure is concentrated on one
dimensional subspace $L_u=\{ z=c u: c \in {\bf R}\}.$
Thus
$$
d \mu_u(v) = e^{-p^2/2\kappa} dp/\sqrt{2 \pi \kappa}, p=(v,w),
$$
where
$
w=\frac{u}{\sqrt{\kappa}}.
$
Thus $||w||=1$. Then
$$
{\bf p}_\mu(j=j_0)=\frac{1}{\kappa} \int_{{\bf R}^n} v_{j_0}^2 d\mu_u
(v)= w_{j_0}^2
$$
This is nothing
else than the {\it Born's rule} that is used in quantum mechanics. The normalized vector
$w$ can be interpreted as a ``pure quantum state.'' In our approach it is nothing else
than the symbolic representation of the Gaussian random vector with the probability
distribution $\mu_u.$ We could do it the other way around -- by starting directly with
an arbitrary normalized vector $w\in {\cal H}_n,$ a ``pure state''. We remark
that Gaussian random vector was chosen only for simplicity. The same result is reproduced
by random vector having the same covariance matrix. Detectors under consideration
(quantum detectors) are not able to distinguish two random vectors with the same covariance
matrix.

We also emphasize that in our approach there is no difference between
``pure'' and ``mixed'' quantum states. All quantum states are simply symbols
for corresponding random vectors.

\subsection{Born's rule}

To write this rule similarly to original Born's formula, we
consider the state space ${\cal H}_n$ as a space of functions $v: X_n \to {\bf
R},$ where $X_n=\{x_1, \ldots, x_n\}$ is some discrete set.
Thus, instead of the set of labels $\{j=1,2,...,n\},$ we now consider
an arbitrary discrete set $X_n$ -- the space of results of measurement.
Any vector $v
\equiv v(x).$ We can say that the measurement process under
consideration is the $X$-measurement: $X=X(v).$ We have
$$
{\bf p}_\mu(X=x)=\frac{1}{\kappa} \int v^2 (x) d\mu (v)
$$
or
$$
{\bf p}_\mu(x \in I)=\frac{1}{\kappa} \sum_{X \in I} \int v^2 (x) d\mu
(v)= \frac{1}{\kappa} \int \sum_{x \in I} v^2 (x) d \mu (v), x \in
X_n, I \subset X_n.$$

In particular, for the Gaussian measure
$\mu_u,$ we have:
$$
{\bf p}_{\mu_u}(X=x)= w^2 (x), \; w=\frac{u}{\sqrt{\kappa}}.
$$

By choosing $n=2k$ and considering the complex  representation of
the real state space ${\cal H}_{2k},$ namely,
 $$
{\cal H}_{2k}={\bf C}^k,
$$
we obtain the theory of measurement in that (as in the previous considerations):
$$
P(X=x|v)=\frac{|v(x)|^2}{||v||^2}, P(v)) = \frac{||v||^2}{\kappa}
\mu (v),
$$
and, finally,
$$
{\bf p}_\mu(X=x)=\frac{1}{\kappa} \int_{{\bf C}^n} |v(x)|^2 d\mu (v),
$$
in particular,
$$
{\bf p}_{\mu_u}(X=x)=|w(x)|^2.
$$

\medskip

{\bf Remark.} We can  consider as the state space any space ${\cal
H}_n= T \times \ldots \times T,$ where $T$ is a number field with
the valuation $|\cdot|_T.$ For example, $T$ can be chosen as the
field of $p$-adic numbers. In this case we consider a probability
$\mu$ on $Q_p^n$ and
$$
P(X=x|v)=\frac{|v(x)|^2_p}{||v||^2}, \; ||v||^2=\sum_{k=1}^n |v(x_k)|^2_p.
$$

\medskip

We emphasize that the theory of measurement under consideration is
purely classical. No noncommutative mathematics was involved in the
description. The crucial point was the use of Bayes' formula (and nothing else!).
From the physical viewpoint, all  "nonclassical features" are induced by special
functioning of
measurement devices ("detectors").

\section{The position measurement for the prequantum field}
\label{ALG}

\subsection{Classical random fields}

We consider the configuration space of  complex random fields: $
Z= L_2 ({\bf R}^3), $ the space of square integrable complex
fields, $\phi: {\bf R}^3 \to {\bf C}.$ It is endowed with the norm
$ \Vert \phi \Vert^2=  \int_{{\bf R}^3}  \vert \phi(x)\vert^2 dx.
$ A random field is  a $Z$-valued random variable $\phi(\omega)
\in Z.$  Here $\omega$ is a chance parameter.  We denote by $\mu$
the probability distribution of the random variable $\phi(\omega)
.$ It is a probability measure on $Z.$\footnote{To be completely
rigorous mathematically, we should consider a Kolmogorov
probability space $(\Omega, F,  {\bf P})$, where $\Omega$ is the
space of chance parameters, $F$ is a $\sigma$-algebra of its
subsets and  ${\bf P}$ is a probability. Then $\phi: \Omega \to Z$
is a measurable function. Its probability distribution is given by
$\mu(U) = {\bf P} (\omega \in \Omega:  \phi(\omega) \in U)$ for a
Borel subset $U$of $Z.$} Of course, $Z$ has infinite dimension,
but mathematical theory of such measures is well developed. We
shall proceed on the physical level of rigorousness. We shall
consider only random fields with zero mean value: $E
\phi(\omega)=0.$ This equality simply means that for any function
$f\in Z:$
 $
 E \int_{{\bf R}^3} f(x) \overline{\phi(x,\omega)} dx=0.
 $
Covariance of a random field is defined as
$$
C_\mu(f, g)= E  \Big(\int_{{\bf R}^3} f(x)
\overline{\phi(x,\omega)} dx\Big) \Big(\int_{{\bf R}^3}
\overline{g(x)}\phi(x,\omega) dx\Big)
$$
The corresponding operator is denoted by $C_\mu.$ We recall that
the covariance operator has all main features of von Neumann's
density operator (self-adjoint, positively defined, trace class),
besides of normalization of the trace by 1. Dispersion is given by
$ \sigma^2(\mu)= E \Vert \phi(\omega) \Vert^2. $ We shall consider
dispersion as a parameter, say $\kappa,$ of our model. We remind
the following useful equality: $ \kappa= \rm{Tr} \; C_\mu.
$

For each $\omega_0,$ the realization of a random field
$\phi(\omega_0)$ is an $L_2$-function. Thus a random field can be
written as a function of two variables $\phi(x,\omega),  x \in
{\bf R}^3.$

\subsection{Random field-signals and position measurement}

 We denote by ${\it E}_\mu$ an ensemble of fields  represented
by a probability measure
$\mu$ on $Z.$  It is the probability distribution of a random
field $\phi(\omega).$ The ${\it E}_\mu$ gives realizations of the
corresponding random field. By our model each ``quantum particle''
is just a symbolic representation of ``prequantum random field''.

Let $X$ be the position observable. This observable is considered
as  an observable on fields. Thus $ X(\omega) =X(\phi(\omega))
$
is a random variable. It takes its values in ${\bf R}.$ Our aim is
to find the probability distribution of this random variable from
the probability distribution of the random field.

In accordance with measurement theory a source of identically prepared
quantum particles is given. In our model this source produces a sequence
of random fields: $\phi^{(1)}(x,\omega), ...., \phi^{(N)}(x,\omega).$
By performing $X$-measurements
for this sequence of random fields we obtain its probability distribution.

In measurement theory for position we consider  a
random field as a {\it random signal} interacting with detectors.
Suppose that the position observable $X$ is given by a measurement
device $M_X.$ For example, $M_X$ can be chosen as a collection of
detectors located at all points $x \in {\bf R}^3.$ For any point
$x_0,$ we can consider the observable $X_{x_0}$ given by a
detector $M_X(x_0)$  located at $x_0.$ For any (sufficiently
regular) set $I \subset {\bf R}^3$ we can consider the observable
$X_{I}$ given by a collection of detectors $M_X(I)$ located  in
the domain $I \subset {\bf R}^3.$

\section{Power signal-field detection theory -- PFSDT}

In our model of detection the measurement process over a random
field consists of two steps:

\medskip

a) selection of a field $\phi \in {\it E}_\mu;$

b) measurement on this field: $X(\phi).$

\medskip

We assume that measurement devices (detectors) are sensitive  to
the {\it power}  of the (classical) field-signal. As in classical
signal theory, we define the power of the field-signal $\phi$ at
the point $x_0$ as
$$
\pi_2 (x_0, \phi)=|\phi(x_0)|^2,
$$
the field-signal power in the domain  $I \subset {\bf R}^3$ is defined
as $ \pi_2 (I, \phi)= \int_I |\phi(x)|^2 dx, $ and finally, the
total power of the field-signal $\phi$ is given by $ \pi_2(\phi)=
\Vert \phi \Vert^2= \int_{{\bf R}^3}|\phi (x)|^2 dx.
$
We now formulate the fundamental feature of the class of detectors
under consideration, namely, {\it sensitivity to the power of a
field-signal} in the form of two postulates:

\medskip

{\bf Postulate 1.} {\it The probability $P_\mu$ to select a fixed
field $\phi$ from the random field-signal $\phi(x, \omega)$ (the
ensemble ${\it E}_\mu$) is proportional to the total power of
$\phi:$}
\begin{equation}
\label{X} dP_\mu (\phi)=K_\mu \pi_2(\phi) d\mu(\phi).
\end{equation}

The coefficient of proportionality $K_\mu$ can be found from the
normalization of probability by one:
$
K_\mu=\frac{1}{\int_Z\pi_2(\phi) d\mu (\phi)}=
\frac{1}{\sigma^2(\mu)}= \frac{1}{\kappa}. $ Thus, we get
\begin{equation}
\label{X1} d P_\mu (\phi)=\frac{\pi_2(\phi)}{\kappa} d\mu (\phi),
\end{equation}
and, for any Borel subset $U \subset Z,$ we have
\begin{equation}
\label{X2} P_\mu (\phi \in U)=\frac{1}{\kappa} \int_U ||\phi||^2
d\mu (\phi),
\end{equation}
or in the random field notations:
\begin{equation}
\label{X3} P_\mu(\phi \in U)=\frac{1}{\kappa} E \Big(\chi_U
(\phi(\omega))||\phi(\omega)||^2\Big),
\end{equation}
where $\chi_U (\phi)$ is the characteristic function of the set
$U.$

 The selection procedure of a signal from a random field for the position measurement
was formalized by Postulate 1. This postulate is  intuitively
attractive: more powerful signals are selected more often.

We now formalize the b-step of the $X$-measurement in the
following form:

\medskip

{\bf Postulate 2.} {\it{The probability $P(X=x_0|\phi)$ to get the
result $X=x_0$ for the fixed field $\phi$ is proportional to the
power $\pi_2(x_0, \phi)$ of this field at $x_0.$ The coefficient
of proportion does not depend on $x_0, $ so $k(x_0\vert \phi)
\equiv k_\phi.$}}

\medskip

The coefficient of proportion  $k(x_0|\phi)$ can be obtained from
the normalization of probability by one:
$
1=\int_{{\bf R}^3} P(X=x|\phi)dx= k_\phi \int_{{\bf R}^3}|\phi(x)|^2 dx. $
Thus
\begin{equation}
\label{CP} k_\phi=\frac{1}{||\phi||^2}.
\end{equation}

The probability to get $X=x_0, x_0 \in {\bf R}^3,$ for a random field
with the probability distribution  $\mu$ can be obtained by using
the classical Bayes' formula: $$ {\bf p}_\mu (X=x_0)= \int_Z P
(x_0|\phi) d P_\mu(\phi)
$$
\begin{equation}
\label{XP1} =  \int_Z \frac{|\phi(x_0)|^2}{||\phi||^2} d
P_\mu(\phi).
\end{equation}
Thus, finally, we have:
\begin{equation}
\label{XP} {\bf p}_\mu  (X=x_0)=\frac{1}{\kappa} \int_Z
|\phi(x_0)|^2 d\mu(\phi).
\end{equation}
Of course, ${\bf p}_\mu  (X=x)$ should be considered as the
density of probability: $$ {\bf p}_\mu (X \in I)=\int_I {\bf
p}_\mu (X=x) dx
$$
\begin{equation}
\label{XP2}
=\frac{1}{\kappa} \int_{Z} \Big(\int_I|\phi(x)|^2
dx\Big) d\mu (\phi),
\end{equation}
where $I$ is  a Borel subset of ${\bf R}^3,$ e.g. a cube.

\section{ Coupling between  PFSDT and QM}

To  find coupling between  PFSDT and the quantum formalism,
 we introduce projectors $\hat{I} \phi (x)=\chi_I (x) \phi (x),$ where $\chi_I$ is the
characteristic function of the Borel set $I.$

\medskip

{\bf Theorem 1.}  {\it The probability measure ${\bf p}_\mu $ can
be represented in the following operator form:
\begin{equation}
\label{QP2} {\bf p}_\mu  (I)=\rm{Tr} \; \hat{\rho}\;  \hat{I},
\end{equation}
where}
\begin{equation}
\label{HY} \hat{\rho} = C_\mu/\kappa.
\end{equation}

We remark that the operator $\hat{\rho}$ has all properties of the
von Neumann density operator. This theorem motivates the following
correspondence between classical random fields and von Neumann's
density operators (quantum states):

\medskip

{\it  Any classical random field induces a quantum state by
mapping the field in the density operator given by (\ref{HY}) and
vice versa.}

\medskip

The probability distribution ${\bf p}_\mu $ on ${\bf R}^3$ given
by PFSDT coincides with the probability distribution given by QM.

We remark that the correspondence between random fields and
quantum states is not one-to-one: a random field is not determined
uniquely by its covariance operator. However, if one restricts
considerations to only {\it Gaussian random fields,} then the
correspondence will become one-to-one.

Let $\Psi$ be a normalized vector --  a "pure state" of QM.  We
consider a measure $\mu_\Psi$ on $Z$ with zero mean value and
covariance operator: $C_\Psi= \Psi \otimes \Psi.$ In particular,
we can choose a Gaussian measure.
 Then we find easily that
\begin{equation}
\label{QR} {\bf p}_\Psi(x)\equiv {\bf
p}_{\mu_\Psi}(x)=|\Psi(x)|^2.
\end{equation}
Thus
\begin{equation}
\label{QP1} {\bf p}_\Psi(I)=\int_I |\Psi(x)|^2 dx.
\end{equation}
This is nothing else than {\it Born's rule.}

We find the mean value of the position $x$ with respect to  the
probability measure ${\bf p}_\mu .$ We restrict our considerations
to one dimensional case. We have:
$$
<x>_{{\bf p}_\mu } =\int_{- \infty}^{+ \infty} x d{\bf p}_\mu (x)
=\rm{Tr} \; \hat{\rho}\;  \hat{x},
$$
where $\hat{x}$ is the position operator in Scr\"odinger's
representation of QM.

Thus  PFSDT (measurement model for classical random fields) produces
the same probability distributions and averages as the
conventional QM model.  As we will see, PFSDT provides a possibility to go beyond
QM.

In appendix, we present the general scheme of representation of quantum
measurement as measurements on prequantum random fields.

\section{Prequantum classical statistical field theory -- PCSFT}

We define {\it ``classical statistical models''} in the following
way: a) physical states $\phi$ are represented by points of some
set $Y$ (state space); b) physical variables are represented by
functions $f: Y \to {\bf R}$ belonging to some functional space
$V(Y);$ c) statistical states are represented by probability
measures on $Y$ belonging to some class $S(Y);$ d) the average of
a physical variable (which is represented by a function $f \in
V(Y))$ with respect to a statistical state (which is represented
by a probability measure $\mu \in S(Y))$ is given by
\begin{equation}
\label{AV0} < f >_\mu \equiv \int_Y f(\phi) d \mu(\phi) .
\end{equation}
A {\it classical statistical model} is a pair $M=(S, V).$

We also recall the definition of the conventional quantum
statistical model with the complex Hilbert state space $H_c.$ It
is described in the following way: a) physical observables are
represented by operators $\hat{A}: H_c \to H_c$ belonging to the
class of continuous self-adjoint operators ${\it L}_s \equiv {\it
L}_s (H_c);$ b) statistical states are represented by von Neumann
density operators (the class of such operators is denoted by ${\it
D} \equiv {\it D} (H_c));$ d) the average of a physical observable
(which is represented by the operator $\hat{A} \in {\it L}_s
(H_c))$ with respect to a statistical state (which is represented
  by the density operator $\hat{\rho} \in {\it D} (H_c))$ is given by von Neumann's
formula:
\begin{equation}
\label{AV1} <\hat{A} >_D \equiv \rm{Tr}\; \hat{\rho} \hat{A}
\end{equation}
The {\it quantum statistical model} is the pair $N_{\rm{quant}}
=({\it D}, {\it L}_s).$

\medskip

We are looking for a classical statistical model $M=(S, V)$ which
will provide {\it ``dequantization'' of the quantum model}
$N_{\rm{quant}} =({\it D}, {\it L}_s).$ By dequantization we
understand constructing of a classical statistical model such that
averages given by this model can be approximated by quantum
averages. Approximation is based on the asymptotic expansion of
classical averages with respect to a small parameter. The main
term of this expansion coincides with the corresponding quantum
average. Such a classical statistical model can be called a
prequantum model. It is considered as to be more fundamental than
QM. The latter provides only an approximative representation of a
prequantum model. Our aim is prove that a prequantum model exists.

We choose the phase space $Y= Q\times P,$ where $Q=P=H$ and $H$ is
the real (separable) Hilbert space.  We consider $Y$ as the real
Hilbert space with the scalar product $(\phi_1, \phi_2)= (q_1,
q_2) + (p_1, p_2).$ We denote  by $J$ the symplectic operator on
$Y:
 J= \left( \begin{array}{ll}
 0&1\\
 -1&0
 \end{array}
 \right ).$
Let us consider the class ${\it L}_{\rm symp} (Y)$ of bounded
${\bf R}$-linear operators $\hat{A}: Y \to Y$ which commute with
the symplectic operator:\begin{equation} \label{SS} \hat{A} J= J
\hat{A} .
\end{equation}
This is a subalgebra of the algebra of bounded linear operators
${\it L} (Y).$ We also consider the space of ${\it L}_{\rm{symp},
s}(Y)$ consisting of self-adjoint operators.

By using the operator $J$ we can introduce on the phase space $Y$
the complex structure. Here $J$ is realized as $-i.$ We denote $Y$
endowed with this complex structure by $H_c: H_c\equiv Q\oplus i
P.$ We shall use it later.

Let us consider the functional space $V(Y)$ consisting of
functions $f:Y \to {\bf R}$ such that: a) the state of vacuum is
preserved\footnote{The vacuum state is such a classical field
which amplitude is zero at any point $x.$} : $f(0)=0;$ b) $f$ is
$J$-invariant: $f(J\phi)= f(\phi);$ c) $f$ can be extended to the
analytic function having the exponential growth: $ \vert
f(\phi)\vert \leq c_f e^{r_f \Vert \phi \Vert} $ for some $c_f,
r_f \geq 0.$ The latter condition provides the possibility to
integrate such functions with respect to Gaussian measures.

The following mathematical result plays the fundamental role in
establishing classical $\to$ quantum correspondence: {\it Let $f$
be a smooth $J$-invariant function. Then } $f^{\prime
\prime}(0)\in {\it L}_{\rm{symp}, s}(Y).$ In particular, a
quadratic form is $J$-invariant iff it is determined by an
operator belonging to ${\it L}_{\rm{symp}, s}(Y).$

We consider the space statistical states $S^{\kappa}(Y)$
consisting of measures $\mu$ on $Y$ such that: a) $\mu$ has zero
mean value; b) it is a Gaussian measure; c) it is $J$-invariant;
d) its dispersion has the magnitude $\kappa.$ Thus these are
$J$-invariant Gaussian measures such that
$$ \int_Y \phi d\mu(\phi)=0 \; \mbox{and}\; \sigma^2(\mu)=
\int_Y \Vert \phi\Vert^2 d \mu(\phi)= \kappa, \; \kappa \to 0.
$$
Such measures describe small Gaussian fluctuations.

We now consider the complex realization $H_c$ of the phase space
and the corresponding complex scalar product $<\cdot, \cdot>.$ We
remark that the class of operators ${\it L}_{\rm symp} (Y)$ is
mapped onto the class of ${\bf C}$-linear operators ${\it
L}(H_c).$ We also remark that, for any $\hat{A}\in {\it
L}_{\rm{symp}, s}(Y),$ real and complex quadratic forms coincide:
$ (A\psi,\phi) =<A\psi,\phi>.$
We also define for any measure its (complex)covariance operator by
$$ <C_\mu y_1, y_2>=\int <y_1, \phi> <\phi, y_2> d \mu (\phi).$$

 We consider now the one parameter family of classical statistical
models:
\begin{equation}
\label{MH} M^\kappa= ( S^\kappa(Y),V(Y)), \; \kappa\geq 0,
\end{equation}

By making in the Gaussian infinite-dimensional integral the change
of variables (field scaling):
\begin{equation}
\label{SC} \phi \to \phi/ \sqrt{\kappa}
\end{equation}
we obtain the following result \cite{KHR}-- \cite{KHR2}:

\medskip

{\it Let $f \in V(Y)$ and let $\mu \in S^\kappa(Y).$ Then the
following asymptotic equality holds:
\begin{equation}
\label{ANN3} <f>_\mu =  \frac{\kappa}{2} \; \rm{Tr}\; \hat{\rho}
\; f^{\prime \prime}(0) + O(\kappa^2), \; \kappa \to 0,
\end{equation}
where the operator $\hat{\rho}= C_\mu/\kappa.$}

\medskip
We see that the classical average (computed in the model
$M^\kappa= ( S^\kappa(Y),V(Y))$ by using the measure-theoretic
approach) is coupled through (\ref{ANN3}) to the quantum average
(computed in the model $N_{\rm{quant}} =({\it D}(H_c),$ ${\it
L}_{{\rm s}}(H_c))$ by the von Neumann trace-formula).

The equality (\ref{ANN3}) can be used as the motivation for
defining the following classical $\to$ quantum map $T$ from the
classical statistical model $M^\kappa= ( S^\kappa,V)$ onto the
quantum statistical model $N_{\rm{quant}}=({\it D}, {\it L}_{{\rm
s}}):$
\begin{equation}
\label{Q20} T: S^\kappa(Y) \to {\it D}(H_c), \; \;
\hat{\rho}=T(\mu)=\frac{C_\mu}{\kappa}
\end{equation}
(the Gaussian measure $\mu$ is represented by the density matrix
$\hat{\rho}$ which is equal to the covariance operator of this
measure normalized by  $\kappa$);
\begin{equation}
\label{Q30} T: V(Y) \to {\it L}_{{\rm s}}(H_c), \; \; \hat{A}=
T(f)= \frac{1}{2} f^{\prime\prime}(0).
\end{equation}
Our previous considerations can be presented in the following form
\cite{KHR}-- \cite{KHR2}:

\medskip

{\bf Beyond QM Theorem.}  {\it The one parametric family of
classical statistical models $M^\kappa= ( S^\kappa(Y),V(Y))$
provides dequantization of the quantum model $N_{\rm{quant}}
=({\it D}(H_c),$ ${\it L}_{{\rm s}}(H_c))$ through the pair of
maps (\ref{Q20}) and (\ref{Q30}). The classical and quantum
averages are coupled by the asymptotic equality (\ref{ANN3}).}

\section{Deviation from predictions of quantum mechanics}
\label{GGG}

Position measurements of higher precision  arise very naturally by
generalization of power field-signal detection theory -- PFSDT to
match the general PCSFT-framework, i.e., consideration of
prequantum physical variables $f(\phi)$ which are given by
nonquadratic functionals of classical field. The appearance  of
additional terms in (\ref{ANN3}) induces deviations from
predictions of QM for averages. Now we would like to find
corresponding deviations for probabilities of detection, namely,
deviations from {\it Born's rule.}

We restrict our modelling   to the
case of fourth order polynomials of classical fields. The main
point is that, instead of the quadratic power of a field-signal
$\phi$ given by $\pi_2(\phi) = \Vert \phi \Vert^2,$ we shall
consider its perturbation by integral of the fourth ´power of
$\phi(x).$ So, we repeat the PFSDT-scheme for the position
measurement in this framework.

At the moment it is not so easy to provide an adequate physical realization
of theoretical measurement scheme which will provide a possibility to check
our main prediction -- {\it violation of Born's rule.}  One of possibilities is based on creation of
 detectors which will be much more sensible to fluctuations of the
prequantum random field than the present  detectors. However, as
was pointed by one of referees of this paper,  {\small ``since a
noisy detector can be made less noisy by averaging over a larger
ensemble, it would seem that the detector hardware may not need to
be replaced, but just the sensitivity of the experiment in which
the detector is used needs to be high.''} It is an extremely
important point. It seems that we need not wait for new
technological jumps in detectors' development. Already now one can
try to design experiments which will be more sensitive to the
probabilistic structure of QM.

And really such an experiment was recently performed \cite{WWW}
demonstrating (at least preliminary) inconsistency of the
probabilistic structure of QM with experimental data. It can be
considered as the first experimental confirmation of the
predictions of PCSFT, see \cite{KHR}--\cite{KHR2}, \cite{PLA}; see
section \ref{SOR} for discussion.

 \bigskip

The measurement process over a random field again consists of two
steps: a) selection of a field $\phi \in {\it E}_\mu$ from an
ensemble of fields ${\it E}_\mu;$ b) measurement on this field:
$X(\phi).$

\medskip

We assume that measurement devices (detectors) are sensitive  to a
"2+4"-{\it power}  of the (classical) field-signal. We define this
power of the field-signal $\phi$ at the point $x_0$ as
$$
\pi_{2,4}(x_0, \phi)=|\phi(x_0)|^2+ |\phi(x_0)|^4,
$$
the field-signal "2+4"-power in the domain  $I \subset {\bf R}^3$ is
defined as $ \pi_{2,4} (I, \phi)= \int_I (|\phi(x)|^2+
|\phi(x)|^4) dx,
$
and finally, the total "2+4"-power of the field-signal $\phi$ is
given by
$
\pi_{2,4}(\phi)=  \int_{{\bf R}^3} (|\phi (x)|^2 + |\phi (x)|^4) dx. $
We remark that, since dispersion $\kappa$ of  a random field
$\phi(x, \omega)$ is considered as a small parameter of the model
(so statistically the field is concentrated in a neighborhood of
$\phi\equiv 0),$ the additional perturbation term $\int_{{\bf R}^3}
|\phi (x,\omega)|^4$  dx is small from the point of view of random
fluctuations.

We now formulate the fundamental feature of the class of detectors
under consideration, namely, {\it sensitivity to the "2+4"-power
of a field-signal} in the form of two postulates:

\medskip

{\bf Postulate 1: "2+4"-power.} {\it The probability $P_\mu$ to
select a fixed field $\phi$ from the random field-signal $\phi(x,
\omega)$ (the ensemble ${\it E}_\mu$) is proportional to the total
"2+4"-power of $\phi:$}
\begin{equation}
\label{X45} dP_\mu (\phi)=K_\mu \pi_{2,4} (\phi) d\mu(\phi).
\end{equation}

The coefficient of proportionality $K_\mu$ can be found from the
normalization of probability by one:
$
K_\mu=\frac{1}{\int_Z \pi_{2,4}(\phi) d\mu (\phi)}. $ Thus, we get
\begin{equation}
\label{X145} d P_\mu
(\phi)=\frac{\pi_{2,4}(\phi)}{\int_Z\pi_{2,4}(\phi) d\mu (\phi)}
d\mu (\phi),
\end{equation}
and, for any Borel subset $U \subset Z,$ we have
\begin{equation}
\label{X245} P_\mu (\phi \in U)=\frac{1}{\int_Z\pi_{2,4}(\phi) d\mu
(\phi)} \int_U \pi_{2,4}(\phi) d\mu (\phi),
\end{equation}
or in the random field notations:
\begin{equation}
\label{X345} P_\mu(\phi \in U)=\frac{1}{E \pi_{2,4}(\phi(\omega))} E
\Big(\chi_U (\phi(\omega))\pi_{2,4}(\phi(\omega)\Big),
\end{equation}
where $\chi_U (\phi)$ is the characteristic function of the set
$U.$

The selection procedure of a signal from a random field for the
position measurement was formalized by Postulate 1. We now
formalize the b-step of the $X$-measurement in the following form:

\medskip

{\bf Postulate 2: "2+4"-power.} {\it{The probability
$P(X=x_0|\phi)$ to get the result $X=x_0$ for the fixed field
$\phi$ is proportional to the "2+4"power $\pi_{2,4}(x_0, \phi)$ of
this field at the point $x_0.$ The coefficient of proportion does
not depend on $x_0, $ so $k(x_0\vert \phi) \equiv k_\phi.$}}

\medskip

The coefficient of proportion  $k(x_0|\phi)$ can be obtained from
the normalization of probability by one:
$
1=\int_{{\bf R}^3} P(X=x|\phi)dx= k_\phi \int_{{\bf R}^3}\pi_{2,4}(x,\phi)
dx. $ Thus
\begin{equation}
\label{CP45} k_\phi=\frac{1}{\pi_{2,4}(\phi)}.
\end{equation}

The probability to get (for the position observation) the result
$X=x_0, x_0 \in {\bf R}^3,$ for a random field with the probability
distribution $\mu$ can be obtained by using the classical Bayes'
formula: $$ {\bf p}_\mu (X=x_0)= \int_Z P (x_0|\phi) d P_\mu(\phi)
$$
\begin{equation}
\label{XP145}
= \int_Z \frac{|\phi(x_0)|^2
+|\phi(x_0)|^4}{\pi_{2,4}(\phi)} d P_\mu(\phi).
\end{equation}
Thus, finally, we have: $ {\bf p}_\mu
(X=x_0)$
\begin{equation} \label{XP45}
=\frac{1}{\int_Z\pi_{2,4}(\phi) d\mu (\phi)} \int_Z
(|\phi(x_0)|^2 +|\phi(x_0)|^4) d\mu(\phi).
\end{equation}
Of course, ${\bf p}_\mu  (X=x)$ should be considered as the
density of probability: $${\bf p}_\mu (X \in I)=\int_I {\bf p}_\mu
(X=x) dx $$
\begin{equation}
\label{XP245} =\frac{1}{\int_Z\pi_{2,4}(\phi) d\mu (\phi)} \int_{Z}
\Big(\int_I(|\phi(x)|^2+|\phi(x)|^4) dx\Big) d\mu (\phi),
\end{equation}
where $I$ is  a Borel subset of ${\bf R}^3,$ e.g. a cube. It is
convenient to make the field scaling (\ref{SC}) to move from the
probability $\mu$ having dispersion $\kappa$ to the corresponding
normalized probability $\nu.$ By this scaling we find direct
dependence of probabilities of detection on the small parameter
$\kappa.$ Then we can represent the coefficient of proportion as $
K_\mu= \frac{1}{\kappa + \kappa^2 c_4}, $ where $ c_4= \int_Z
\int_{-\infty}^\infty |\phi(x)|^4 dx d\nu(\phi).
$
We find the following dependence on the small parameter $\kappa$
(the dispersion of random fluctuations) of the probability of the
position detection: $$ {\bf p}_\mu  (X=x_0)=\frac{1}{\kappa +
\kappa^2 c_4} \int_Z ( \kappa |\phi(x_0)|^2 +
\kappa^2|\phi(x_0)|^4)d\nu(\phi) $$
\begin{equation}
\label{XPP45} = \frac{1}{1 + \kappa c_4} \int_Z (|\phi(x_0)|^2 +
\kappa|\phi(x_0)|^4) d\nu(\phi).
\end{equation}

If in the formula (\ref{XP}) (for PFSDT with quadratic power)  we
make scaling (\ref{SC}), we obtain:
\begin{equation}
\label{XPPT} {\bf p}_\mu(X=x_0)= \int_Z |\phi(x_0)|^2 d\nu(\phi).
\end{equation}
The same result we obtain by the considering the limit $\kappa\to
0$ in (\ref{XPP45}). Thus the model presented in this section is
really the $O(\kappa)$ perturbation of the model considered  in
section 3 (and hence of QM).

We come back to our model of detection taking into account the
fourth power of the signal-field: $ {\bf p}_\mu  (X\in I) $
$$
=\frac{ \int_Z \int_I |\phi(x)|^2 dx d\nu (\phi) + \kappa\int_Z
\int_I|\phi(x)|^4 dx d\nu(\phi)}{1 + \kappa c_4}.
$$
 Hence
${\bf p}_\mu  (X\in I)$ $$ \approx (1 - \kappa c_4) \Big(\int_Z
\int_I |\phi(x))|^2 dx  d\nu (\phi) + \kappa \int_Z
\int_I|\phi(x)|^4 dx) d\nu (\phi)\Big)
  $$
$$
=\int_Z  \int_I|\phi(x)|^2 dx d\nu (\phi)) + \kappa \Big[ \int_Z
\int_I |\phi(x)|^4 dx d\nu (\phi)
$$
$$
 -\int_Z  \int_I|\phi(x)|^2 dx  d\nu (\phi) \times
\int_Z  \int_{{\bf R}^3}|\phi(x)|^4 dx d\nu (\phi)\Big].
$$
The first summand gives the well known Born's rule (the
conventional QM-prediction). The second summand (we denote it by
$\Delta (I, \mu, \kappa)$) gives the deviation from the Born's
rule.

We consider now this deviation in the case of so called pure state
$\Psi, ||\Psi||=1.$ In our approach this corresponds to the case
when the normalized measure $\nu$ is the Gaussian measure
$\nu\equiv \nu_\Psi$ with the covariance operator:
$C=\Psi\otimes\Psi.$ We have $\Delta (I, \Psi, \kappa)$
\begin{equation}
\label{GI} =\kappa \Big[\int_I|\Psi(x)|^4 dx - \int_I|\Psi(x)|^2
dx \int_{{\bf R}^3}|\Psi(x)|^4 dx\Big]
\end{equation}
 Suppose\footnote{To be mathematically rigorous, we consider $\Psi \in L_{2, 4}({\bf R}^3):$ both integrals
 $\int |\phi(x)|^2 dx$ and $\int |\phi(x)|^4 dx$ are finite.} that $\rm{supp}\; \Psi \subset I,$ so the wave function is zero outside
 the set $I.$ Then $\Delta \equiv 0.$

Further we consider one dimensional case. Let now $\Psi(x)=H, L/2
\leq x \leq L/2.$ Thus $ H^2 L=1,$ so $L=1/H^2.$ We choose $I=[0,
L/2]:$
$$
\int_I|\Psi(x)|^2 dx =1/2, \int_{{\bf R}^3}|\Psi(x)|^4 dx= H^4
L=H^2,
$$
and
$$
\int_I|\Psi(x)|^4 dx=\frac{H^4 L}{2}=\frac{H^2}{2}, \;
\Delta=\kappa (\frac{H^2}{2} - \frac{H^2}{2})=0.
$$
This calculation gave a hint that an asymmetric probability
distribution may induce  nontrivial $\Delta.$ We choose:
\[\Psi(x)= \left\{ \begin{array}{ll}
H, \; -L/2 \leq x \leq 0\\
kH, \; 0 < x \leq L/2
 \end{array}
 \right .
\]
Hence, $1=||\Psi||^2= L H^2 (k^2 + 1)/2,$  so $L= 2/(H^2 (k^2 +
1)).$ Here $I=[L/2, 0], \int_I|\Psi (x)|^2 dx=H^2 L/2= 1/(k^2 +
1);$
$$
\int_{{\bf R}^3}|\Psi(x)|^4 dx=
\Big (\frac{1 + k^4}{1 + k^2} \Big ) H^2, \; \int_I|\Psi(x)|^4
dx=\frac{H^2}{k^2 + 1}.
$$
$$
\Delta=
\frac{\kappa H^2 k^2 (1-k^2)}{(1 + k^2)^2}.
$$
If $k > 1,$ then $\Delta(I, \Psi, \kappa) < 0.$ Suppose that $H$
increases (and $k$ is fixed) then the deviation from Born's rule
will be always negative and this deviation will be increasing. So,
for large $H$, the probability to find a system in $I$ will be
essentially less than predicted by QM. For example, choose $k=2,$
then
$$
\Delta=
 0,48\kappa H^2.
$$
On the other hand, by choosing $k < 1,$ we shall get the positive
deviation. For $k=1,$ we have  $\Delta=0$ and there will be no
deviation from Born's rule.

All our considerations are purely qualitative, since we do not
know the magnitude of $\kappa.$ But one may expect that such a
qualitative effect as decreasing and increasing the probability
(comparing with the Born's rule) can be observed in experiments.

\section{Averages}

We take a nonquadratic functional of classical fields:
$$
f_x(\phi)=\int_{-\infty}^{+\infty} x (|\phi(x)|^2 +|\phi(x)|^4)dx.
$$
It is mapped onto the position operator by the map  $T:$ PCSFT$\to$
QM.  PCSFT gives the following average:
$$
<f_x>_\mu=\int_{-\infty}^{+\infty} x \int_Z(\phi(x)|^2 +
|\phi(x)|^4)d\mu(\phi) dx
$$
$$
= \kappa \int_{-\infty}^{+\infty} x  \int_Z|\phi(x)|^2 d\nu(\phi)
dx $$ $$ + \kappa^2 \int_{-\infty}^{+\infty} x \int_Z|\phi(x)|^4
d\nu(\phi) dx.
$$
On the other hand, by using the distribution provided by PFSDT
(field power detection model) we get:
$$
<x>_{{\bf p}_\mu}=\int_{-\infty}^{+\infty} xd{\bf p}_\mu (x)
$$
$$
= \Big[\kappa\int_{-\infty}^{+\infty} x  \int_Z|\phi(x)|^2 d\nu
(\phi) dx $$ $$+ \kappa^2 \int_{-\infty}^{+\infty} x
\int_Z|\phi(x)|^4 d\nu (\phi) dx\Big]$$ $$/ \Big[\kappa + \kappa^2
\int_{-\infty}^{+\infty} \int_Z |\phi(x)|^4 d \nu (\phi) dx\Big].
$$
We consider the normalization based on the field-power functional
$\pi_{2,4}(\phi).$ Then
\begin{equation}
\label{PW}
\frac{<f>_\mu}{<\pi_{2,4}>_\mu}=\int_{-\infty}^{+\infty} x d{\bf
p}_\mu (x).
\end{equation}
Thus by normalizing the PCSFT-average by such a field-power
functional we obtain the quantity which coincides with
PFSDT-average. The PCSFT-averages should be normalized in the
corresponding way to obtain PFSDT-averages. Of course, this is valid
only for polynomial functionals of the same field-power as
detectors considered by PFSDT.

We now consider the field-power functional $
\pi_2(\phi)=||\phi||^2. $ The basic asymptotic equality of PCSFT can
be written in the form:
$$
\frac{<f>_\mu}{<\pi_2>_\mu}=<T(f)>_{T(\mu)} + O(\kappa), \kappa
\to 0.
$$

\section{PFSDT: scheme for local measurements}
\label{LOCAL}

To obtain a local measurement scheme, we should take into account
that any detector is located in a special domain, say $O,$ of space. More generally,
if a detector measures some variable, e.g., energy, it operates only in a special
range of variation of this variable. This fundamental fact of measurement theory
should be taken into account, cf. Haag \cite{Haag1} and especially
\cite{Haag2},  and our  PFSDT should be modified. We consider
again position measurement. As before, measurement process over a random
field consists of two steps: a) selection of a fixed field $\phi$ from
the random prequantum field; b) measurement on this field.

Thus we proceed under the assumption that the detector operates in the domain
$O.$

\medskip

{\bf Postulate 1.} (Local) {\it Probability to select a fixed
field $\phi$ from the random field-signal $\phi(x, \omega)$
 is proportional to the  power of the field-signal
$\phi$ in the domain $O:$}
$
dP_\mu (\phi\vert O)=K_{\mu,O} \;\pi_2(O; \phi)\; d\mu(\phi).
$

The coefficient of proportionality $K_\mu$ can be found from the
normalization of probability by one:
$
K_{\mu,O} =
\frac{1}{\kappa_O}, $
where $\kappa_O=\int_Z\pi_2(O; \phi) d\mu (\phi).$
Thus, we get
$ d P_\mu (\phi \vert O)=\frac{\pi_2(O; \phi)}{\kappa_O} d\mu (\phi),
$
Postulate 1 is  intuitively
attractive: signals  which are more powerful  in the domain of detection
$O$ are selected more often. We now again formalize the b-step:

\medskip

{\bf Postulate 2.}  (Local) {\it{Probability $P(X=x_0|\phi, O)$ to get the
result $X=x_0,$ where $x_0 \in O,$  for the fixed field $\phi$ is proportional to the
power $\pi_2(x_0, \phi)$ of this field at $x_0.$ The coefficient
of proportion does not depend on $x_0, $ so $k(x_0\vert \phi)
\equiv k_{\phi,O}.$}}

\medskip

This coefficient can be obtained from
the normalization of probability by one:
$
1=\int_{O} P(X=x|\phi,O)dx= k_{\phi, O} \int_{O}|\phi(x)|^2 dx. $
Thus
$k_{\phi,O}=\frac{1}{\pi_2(O; \phi)}.$

The probability to get $X=x_0, x_0 \in O,$ for a random field
with the probability distribution  $\mu$ can be again obtained by using
the classical Bayes' formula: $ {\bf p}_\mu (X=x_0\vert O)= \int_Z P
(x_0|\phi,O) d P_\mu(\phi\vert O)$
$ \int_Z \frac{|\phi(x_0)|^2}{\pi_2(O; \phi)} d
P_\mu(\phi \vert O).$
Thus, finally, we have:
\begin{equation}
\label{XPRRR} {\bf p}_\mu  (X=x_0\vert O)=\frac{1}{\kappa_O} \int_Z
|\phi(x_0)|^2 d\mu(\phi).
\end{equation}
Of course, ${\bf p}_\mu  (X=x\vert O)$ should be considered as the
density of probability. For a domain $I\subset O,$
$$ {\bf p}_\mu (X \in I\vert O)=\int_I {\bf
p}_\mu (X=x\vert O) dx
$$
\begin{equation}
\label{XP2RRR}
=\frac{1}{\kappa_O} \int_{Z} \Big(\int_I|\phi(x)|^2
dx\Big) d\mu (\phi).
\end{equation}

\section{Double clicks problem}
\label{DCL}
 One of the referees presented the following important
objection to the presented model, PFSDT, of detection for PCSFT:

\medskip

{\small ``As there are no point particles explicitly built into
this model, the issue of locality is potentially a concern. For
example, consider two detectors $A$ and $B$ located in volumes
$O_A$ and $O_B$ which are sufficiently far from each other and
consider a single particle wave function whose support
simultaneously includes the volumes of both detectors at the time
of measurement as the wave function moves through both detectors.
Assume ideal detectors so that their quantum efficiency is $100\%$
and there are no false counts. According to quantum mechanics, if
one detector finds a particle then the other detector can not find
one. The conditional probability that a particle is found at $B$
given that one was found at $A$ must vanish. This is a statement
of locality together with the property of a particle. A single
particle can not be detected in two two spacelike separated
4-volumes. This gedanken experiment would seem to cause a problem
for a classical field model. I would expect the classical field
for a pure state also to be non-zero in the volumes of both
detectors. So the probability of detecting simultaneously a
particle at detector $A$ and at detector $B$ as predicted by the
classical field model would be non-zero in contradiction to
quantum mechanics. I think the author should address this issue.
Perhaps the current model is limited to the behavior of a single
detector or to mutiple detectors in high flux beams, but is not
applicable to multiple detectors with single particle quantum
states (or very low flux beams). This would mean that it is a
semiclassical model with a certain domain of applicability, but it
is not completely equivalent to quantum mechanics.''}

\medskip

Of course, the model presented in this paper was elaborated for
only one detector measurement. The problem of its extension to
measurements performed simultaneously by a few detectors has not
been yet solved, including, generalization to multi-particle
systems -- prequantum random fields representing entangled quantum
particles. However, even in the presented framework one can see
that the problem mentioned by the referee could not be
 ignored.
 The simplest solution is to agree (at leat5 at the moment)
 with the referee and consider   PFSDT as a semiclassical model,
i.e., without to pretend to cover QM  completely. It might be  a
right decision, especially since, as it was remarked,
``multi-particle systems'' (e.g., random fields corresponding to
entangled systems) have not been handled. Nevertheless, I would
like to point to some similarity of this problem with the problem
of experimental testing of violation of Bell's inequality. In the
latter case any local model with hidden variables evidently
confronts with the {\it theoretical  formalism of QM.} However,
the experimental situation is extremely complicated. Various
loopholes do not give a possibility to be completely sure that,
e.g., Aspect's experiment \cite{AST} or Weihs-Zeilinger's
experiment \cite{WSP}, \cite{WSP1} provided the final confirmation
of Bell's argument, see \cite{AKK}--\cite{AKK5} for debates.

I suspect that similar loopholes will appear in the experimental
test  considered by the referee. Unfortunately, I do not know so
well the experimental situation in this domain (at least comparing
with Bell's test). My first reflection  is that it is not easy to
prepare a single system (which is sufficiently massive) in e.g.
{\it camel type state} -- with two domains of essential
concentration of the wave function which are sufficiently far from
each other (comparing with the velocity of light). \footnote{We
remind that in PCSFT  pure quantum states are just labels for
classical prequantum random fields. However, the support of a
``quantum wave function'' coincides with the support of
realizations of the corresponding prequantum random field.}  If it
is really the case then one should work with the camel-like states
of photons. However, problems of Bell's test will also appear in
this situation, e.g., the problem of efficiency of detectors or
the time window problem, cf. \cite{WSP1}.

By using loopholes one may speculate that  the ``single detector
click'' prediction of QM is also an idealization which might be
violated in better experiments. However, at the moment it is too
early to present such a conjecture in the PFSDT/PCSFT-framework,
especially in the absence of a few detectors generalization of the
presented measurement theory.  I just remark that the famous
experiment of Grangier \cite{Grangier1},  \cite{Grangier} testing
the semiclassical model can be easily objected on the basis of the
detection loophole (in the same way as Aspect's type experiments).
Moreover, one can easily present a model with hidden variables
which reproduce Grangier's experimental data \cite{EG}, \cite{EG1}
(e.g., by playing with discrimination thresholds).

\section{Discussion on the triple slit experiment}
\label{SOR}

 We remark that the original motivation for
experiment done in \cite{WWW} was based on  Sorkin's works on the
{\it ``sum over histories''} approach to QM, \cite{Sorkin}. The
latter is also a kind of prequantum model. In Sorkin's approach
quantum probabilities are embedded in theory of generalized
``probabilities'' (the latter have  unusual properties, it seems
that the standard frequency interpretation is impossible).
Sorkin's model is essentially more general than QM. In principle,
a model which matches with QM for the two slit interference, but
not for the triple slit can be considered. In such a model Born's
rule should be violated, otherwise there would be no difference
even in the case of the triple slit experiment. This motivated
Sorkin to propose a test for violation of Born's rule based on the
triple slit experiment, see \cite{WWW}.

PSCFT differs crucially from Sorkin's model \cite{Sorkin}. It is
based on ordinary (Kolmogorov) ensemble probability. Thus we need
not appeal to generalized probabilistic models to get quantum
probabilities, cf. Sorkin \cite{Sorkin} and  also Khrennikov
\cite{KKH}. Moreover,  by PCSFT there are no reasons to expect
violation of Born's rule only in the triple (or more) slit
experiment. In principle, violation is expected already in the two
slit experiment. In one of future papers a comparative analysis of
two and triple slit experiments from the point of view of PCSFT
will be performed. Such an analysis may be useful to find
experimental designs which are most sensitive to perturbation of
Born's rule (from the viewpoint of PCSFT).

However, even without such an analysis it is clear that the triple
slit experiment can be considered as increasing of precision of
the position measurement comparing with the two slit experiment.
In accordance with our model higher precision measurement of
position may produce deviation from Born's rule. From this
viewpoint the $n$-slit experiment may produce more visible
deviation for large $n.$

\section{Other tests of Born's rule}

Let $a=\pm 1$ and $b= \pm 1$ be two observables which are
presented in QM by  two self-adjoint operators, $\widehat{a},
\widehat{b}.$ In the case of nondegenerate spectra,they are
realized in the two dimensional complex space. Consider their
eigenvectors $\widehat{a} e_{\pm}^a = \pm e_{\pm}^a, \widehat{b}
e_{\pm}^b = \pm e_{\pm}^b.$

\subsection{Double stochasticity test}

Consider the matrix of transition probabilities $P^{b \vert a}:$
$$
p_{\beta \vert \alpha}\equiv {\bf P}(b=\beta \vert a=\alpha),
$$
where $\alpha, \beta= \pm 1.$

 Both in classical and quantum probabilistic models this
matrix is always {\it left stochastic.} A left stochastic matrix
is a square matrix whose columns consist of nonnegative real
numbers whose sum is 1. Really, for each value $a=\alpha,$
$$
\sum_\beta{\bf P}(b=\beta \vert a=\alpha)=1.
$$However, in QM, unlike the
classical model, this matrix is {\it always} doubly stochastic.

We remind that in a {\it  doubly stochastic matrix} all entries
are nonnegative and all rows and all columns sum to 1: for each
value $b=\beta,$
\begin{equation}
\label{DSDS}
 \sum_\alpha{\bf P}(b=\beta \vert a=\alpha)=1.
\end{equation}
It is a simple consequence of Born's rule, see \cite{INT} for
details.

Any {\it experimental test of double stochasticity} can be
considered as a test of validity of Born's rule. If one finds two
obseravbles such that (\ref{DSDS}) is violated, then the formalism
of QM, in particular, Born's rule cannot be applied.

We remark that, unlike Sorkin's test, the aim of the double
stochasticity test is not to distinguish QM from a more general
nonclassical model. In the classical model the matrix of
transition probabilities can be doubly as well non-doubly
stochastic.  One of consequences of this remark is that the double
stochasticity test is most useful for massive particles (since
double stochasticity takes place for classical electromagnetic
field, e.g., in classical optics).

I propose to check carefully double stochasticity for spin
projections on two axes $a$ and $b,$ e.g., for electrons. First
the spin projection on the axis $a$ is measured in the form of
nondestructive filtration. After the Stern-Gerlach magnet with the
$a$-orientation, electrons with  spins up ($a=+1)$ and down
($a=-1)$ go to different directions. The second Stern-Gerlach
magnet (with  the $b$-orientation) is used for the final
measurement. Transition probabilities are calculated, They are
summed up with respect to $\alpha= \pm$ for each $b=\beta.$

\subsection{Interference magnitude test}

This test is more complicated, since it involves preparation of a
variety of states, i.e., it should be repeated for ensembles of
``particles'' in different states. Consider again two observables
$a= \pm 1$ and $b= \pm 1.$ Take any pure state $\psi.$ Then it is
easy to derive the following {\it formula of interference of
probabilities}, see \cite{INT}:
\begin{equation}
\label{QS7} {\bf P}_\psi (b= \beta)= \sum_{\alpha} {\bf
P}_\psi(a=\alpha) {\bf P} (b=\beta\vert  a=\alpha)
\end{equation}
$$
 + 2 \cos \theta \sqrt{  {\bf P}_\psi(a=\alpha_1) {\bf P}
(b=\beta\vert a=\alpha_1) {\bf P}_\psi(a=\alpha_2) {\bf P}
(b=\beta\vert  a=\alpha_2)},
$$
where $\alpha_1 =+1, \alpha_2= -1.$

Here ${\bf P}_\psi (b= \beta)= \vert \langle \psi, e_\beta^b
\rangle \vert^2, {\bf P}_\psi(a=\alpha)=\vert \langle \psi,
e_\alpha^a \rangle \vert^2, {\bf P}(b=\beta \vert a=\alpha_1)=
\vert \langle e_\beta^b, e_\alpha^a \rangle\vert^2.$ The phase
$\theta$ can be found from phases of considered scalar products.
We remark that (\ref{QS7}) implies the following inequality
\begin{equation}
\label{QS7U}  \Big\vert \frac{{\bf P}_\psi (b= \beta)-
\sum_{\alpha} {\bf P}_\psi(a=\alpha) {\bf P} (b=\beta\vert
a=\alpha)}{2\sqrt{{\bf P}_\psi(a=\alpha_1) {\bf P} (b=\beta\vert
a=\alpha_1) {\bf P}_\psi(a=\alpha_2) {\bf P} (b=\beta\vert
a=\alpha_2)}} \Big\vert \leq 1.
\end{equation}
Since all probabilities involved in this inequality can be
estimated by experimental frequencies, this condition can be
checked experimentally. Any statistically essential deviation from
this inequality can be considered as a sign of violation of Born's
rule (which was basic in the derivation of the formula of
interference of probabilities). Of course, the main problem is to
find an appropriative pair of observables and a preparation
procedure for systems (``pure state'' $\psi.)$

Consider e.g. the case such that all transition probabilities are
equal to 1/2. Set $p_+ ={\bf P}_\psi (b= +)$ and $q_+= {\bf
P}_\psi (a =+).$ Then (\ref{QS7U}) is simplified
\begin{equation}
\label{QS7UU} \vert p_+ - 0.5 \vert > \sqrt{q_+(1-q_+)}.
\end{equation}
Thus both variables should be essentially nonsymmetrically
distributed.

\section{Appendix: Measurement of observables with discrete spectra}

Let $f(\phi)$ be a classical physical variable (functional of
classical fields). Consider corresponding operator $A,$ see
(\ref{Q30}).  By our model quantum observable represented by the operator $A$
performs an approximative measurement of the classical variable
$f.$ For simplicity, suppose that $A$ has purely
discrete spectrum. By our prequantum model detectors described by the formalism of
QM work in the following way:

\medskip
We define the $A$-power of the field-signal $\phi(x)$ at the point
$a \in {\bf R}$ as $\pi_2^{(A)} (a, \phi)=\vert <\phi, e_a>|^2,$
where $A e_a= a e_a;$ on the interval $I\subset {\bf R}$
as $\pi_2^{(A)}(I, \phi)=\sum_{a \in I} \pi_2^A (a, \phi)$
and the total power as
$\pi_2^{(A)}(\phi)=\sum_{a \in {\bf R}} \pi_2^A (a, \phi) \equiv ||\phi||^2.$
We repeat shortly our scheme of measurement based on interaction of an $A$-detector
with a prequantum classical field (paying the role of signal):

a) selection of the  concrete field signal $\phi$ with probability proportional to the
square of norm;

b) production of some value $A=a$ with probability
proportional to the $A$-power of the prequantum field-signal
$\phi$ at the point $a.$

As in previous considerations, these postulates and the conventional
Bayes' formula imply:

${\bf p}_\mu^{(A)} (a)= \frac{1}{\kappa} \int_Z \pi_2^{(A)} (a, \phi)
d\mu (\phi)= \frac{1}{\kappa} \int_Z |<\phi, e_j>|^2 d\mu (\phi).$

Let now $\mu=\mu_\Psi,
||\Psi||=1.$ It is Gaussian measure with zero mean value and covariance
$C= \Psi \otimes \Psi.$  Then $\int_Z|<\phi, e_a>|^2 d\mu_\Psi (\phi)=
|<\Psi, e_a>|^2.$ Hence ${\bf p}_{\mu_\Psi}^{(A)}(a)= |<\Psi,
e_a>|^2.$ This is Born's rule.

A local modification of this model -- as in section \ref{LOCAL} -- is evident.

\section{Appendix: Wiener-Siegel differential
space theory}

There is some similarity between PFSDT/PCSFT theory and the
differential space theory of Wiener and Siegel \cite{WS},
\cite{WS1} which was made more palatable for physicists by Bohm
and Bub (WSBB, \cite{BU}-- \cite{BU2}). Both theories have a
random classical field. In WSBB, there is actually a random
Hilbert space vector, but in the position basis this becomes a
random field. WSBB uses a universal algorithm for finding the
value of a quantum observable given the quantum state vector and
the particular instance of the classical field, called the
polychotomic algorithm. This avoids the locality problem mentioned
in section \ref{DCL}, because it ensures that the particle (for a
single particle wave function) can only be found at one point at
one time. But the polychotomic algorithm is not local to one
detector, but rather selects a position coordinate for the
particle using the whole wave function and the whole classical
random field.

We remark that the PFSDT algorithm presented in section \ref{ALG}
is neither local, since at the first step of its realization a
detector takes into account the total power of the classical field
-- its $L_2$-norm (or integrals with higher nonlinearities in
more general framework). However, the algorithm is easily
transformed into a local detection algorithm, see section
\ref{LOCAL} by using Haag's approach. However, the problem of
double clicks arises.

More detailed comparing of PFSDT/PCSFT with WSBB will be presented
in one of future publications.

\medskip

{\bf Conclusion.} {\it A model of measurement in which ``quantum particles'' are represented
by classical random fields interacting with detectors is presented. The main prediction of this
model is deviation from the basic probabilistic law of QM, Born's rule.
The latter is only an approximative formula.
We expect that numerous experiments demonstrating violation
of Born's rule will be designed. Our prediction has been already confirmed (at least preliminary)
in the experiment \cite{WWW}.}

\end{document}